\documentclass[nofootinbib,
prd,
]{revtex4}

\usepackage{graphicx}
\usepackage{mathrsfs}
\usepackage{yfonts}
\usepackage{tipa}
\usepackage{bm}
\usepackage{bbm}
\usepackage{amsmath}
\usepackage{amssymb}
\usepackage{amsthm}
\usepackage{hyperref}
\usepackage{amsfonts}
\usepackage{mathtools}
\usepackage{latexsym}
\usepackage[greek,english]{babel}
\usepackage{booktabs}
\usepackage[iso-8859-7]{inputenc}

\begin{document}

\title{Regular black holes in isothermal cavity}

\author{Athanasios G. Tzikas}
\email{tzikas@fias.uni-frankfurt.de}
\affiliation{Frankfurt Institute for Advanced Studies (FIAS), Ruth-Moufang-Str. 1, 60438 Frankfurt am Main, Germany}
\affiliation{Institut f\" ur Theoretische Physik, Goethe Universit\" at Frankfurt, Max-von-Laue-Str.1, 60438 Frankfurt am Main, Germany}

\begin{abstract}
We examine the thermodynamic behavior of a  static neutral regular (non-singular) black hole  enclosed in a finite isothermal cavity. The cavity enclosure  helps us investigate black hole systems in a canonical or a grand canonical ensemble. Here we demonstrate the derivation of the reduced action for the general metric of a regular black hole in a cavity by considering a canonical ensemble. The new expression of the action contains quantum corrections at short distances and concludes  to the action of  a singular black hole in a cavity at large distances. We apply this formalism to the noncommutative  Schwarzschild black hole, in order to study the phase structure of the system. We conclude to a possible small/large stable regular black hole  transition inside the cavity that  exists neither at the system of a classical Schwarzschild black hole in a cavity, nor at the asymptotically flat regular black hole  without the cavity. This phase transition seems to be similar with the liquid/gas transition of a Van der Waals gas. 
\end{abstract}

\maketitle

\section{Introduction}
\label{intro}

Black holes are probably among the most fascinating topics in Physics. They are not only mysterious astrophysical objects but also a theoretical laboratory where one can study fundamental physics. 
Indeed, at their center general relativity brings about its own downfall by predicting singularities. This issue can be overcome by fully quantizing  gravity.
However, the already existing candidates of quantum gravity are not helping that much to overcome such a problem. As a result, a literature of effective regular and ultraviolet complete black holes has been proposed
\cite{NSS06,Ans08,Nic09,NiW11,IMN13,NiS14,CMN15,FKN16,Nic18,NSW19,KKNM19},  replacing the curvature singularity of the black hole with a regular  core, while recovering general relativity at large distances. 

On the top of that, a variety of thermodynamic phenomena have been noticed for black holes 
during the years following the derivation of the black hole temperature \cite{Haw74} and the related thermodynamics \cite{Bek73}. Of particular interest is the discovery of the Hawking-Page transition \cite{HaP83} between  pure thermal radiation and a stable anti-de Sitter (AdS) black hole. 
Moreover, charged AdS black holes have an intriguing phase structure \cite{CEJ+99,CEJM99}, in contrast to the uncharged ones, which is recently identified  with the phase  structure of a Van der Waals gas \cite{KuM12}. This result is possible if we consider an extended phase space for the black hole, where the cosmological constant is treated as the thermodynamic pressure of the system \cite{KRT09,KuM17}. Similar phase structures have also been noticed for the known Bardeen black hole in AdS spacetime \cite{Tzi19}.
Another interesting issue is the enclosure of classical black hole systems, uncharged \cite{Akb04,WWY+20}  or charged \cite{BBW+90,PeL99}, inside an isothermal cavity  that acts  as a reservoir. This can be realized by placing a heat bath around the cavity that  fixes the temperature  at its surface, allowing for thermodynamically stable black holes to exist. Such a system is in analogy with the known canonical ensemble in statistical mechanincs. The cavity enclosure appears similarities with the properties of AdS spacetime, which also stabilizes the black hole by acting as a reflecting box. For instance, Carlip and Vaidya \cite{CaV03} found  the same phase structure between pure AdS black holes and asymptotically flat cases in a cavity, as well as de Sitter cases. More recently, new phase transitions and critical phenomena have been discovered for both neutral and charged de Sitter black holes with a Born-Infeld gauge field  inside an extended phase space   \cite{SiM18,SiM19}.

So far, a model of a regular black hole surrounded by an isothermal cavity has not been discussed in the scientific literature. For this reason, it is  the goal of this paper  to  study the thermodynamics  and the phase structure of such a system. The structure of the paper is as follows: In Sec.~\ref{sec:2} we choose as our regular black hole candidate the noncommutative Schwarzschild black hole \cite{NSS06}, which is a non-singular black hole solution of the Einstein equations, and we review its basic geometrical and thermodynamic features. In addition, we provide the basic boundary conditions/constraints for the general system \textit{black hole-cavity}.  In Sec.~\ref{sec:3} we assume the cavity enclosure for any generic regular black hole solution,  resulting from the existence of an anisotropic fluid with two pressure terms,   and we perform the derivation of the reduced action. The new action is a generalization of the action characterizing the classical system \textit{Schwarzchild black hole-cavity} \cite{Lun08} (see also the uncharged version  in \cite{BBW+90}), containing extra quantum correction terms at short scales, due to the existence of the minimal cut-off length that regularizes the spacetime geometry at the center. The action of the regular case concludes to the singular case at the large distance regime. In Sec.~\ref{sec:4} we  consider the model of  the noncommutative Schwarzschild black hole  and we examine the phase structure of the system  \textit{noncommutative black hole-cavity}. In contrast to the classical Schwarzchild black hole,  which appears a Hawking-Page-like phase transition inside the cavity \cite{Lun08}, the regular black hole undergoes a phase transition between a small and a large stable black hole similar to the liquid/gas transition of a Van der Waals gas. Similar behavior has also been noticed for noncommutative \cite{NiT11}  and Reissner-N\"ordstrom \cite{KuM12} black holes inside the AdS space. Here the role of the negative cosmological constant has been replaced by the cavity, indicating  that the cavity acts  to the system the same way as the AdS space does.
Sec.~\ref{sec:concl} is devoted to conclusions.  Throughout the paper we use Planck units ($c=\hbar=G=k_{\mathrm{B}}=1$) and we consider a canonical-like ensemble, where   the boundary radius and the temperature of the cavity wall are held fixed  and the only variable which is allowed to fluctuate is the event horizon.

\section{Basic Review}
\label{sec:2}

\subsection{Noncommutative Schwarzschild black hole}

A basic feature of noncommutative geometry \cite{Nic09} is the presence of a minimal cut-off length that makes gravity ultraviolet self-complete below this scale. The meaning of a smooth classical manifold breaks down when approaching this length because quantum fluctuations are expected to appear, making  spacetime  "fuzzy"  at short distances. These fluctuations can be encoded in the commutator 
\begin{equation}
\left[ \textbf{x}^{\mu}\,, \textbf{x}^{\nu}  \right]=i \theta^{\mu\nu} \,,
\end{equation}
where $\theta^{\mu\nu}$ is an anti-symmetric tensor that discretizes the spacetime vectors $\textbf{x}^{\mu} $ in the same way the Planck constant discretizes the phase space.  The quantity $\sqrt{\theta}$ has dimensions of length and  can be interpreted as the minimal cut-off length, very close to the Planck length, i.e., $\sqrt{\theta} \sim 10^{-33}  \mathrm{cm}\,$, below which there is no physically meaningful length.

It can be shown that noncommutativity implies the replacement of point-like structures, that are described by Dirac-delta functions, with very narrow Gaussian distributions \cite{SmSp03,SSN06++,KoN10}. This feature has been used to improve the Schwarzschild singularity \cite{NSS06} 
by imposing  that the black hole mass is diffused throughout the center, just like a Gaussian distribution of minimal width $\sqrt{\theta}\,$, instead of being a vacuum Dirac-delta function. The  mass density of the anisotropic fluid-source is chosen to be then
\begin{equation} \label{Cncdensity}
\rho(r) = \frac{M}{\left( 4 \pi \theta \right)^{3/2} } \exp (-r^2/4\theta) \,.
\end{equation}
We make the following ansatz for the line element
\begin{equation}
\mathrm{d} s^2 = -b^2(r) \mathrm{d} t^2 + a^2(r) \mathrm{d} r^2 + r^2 \mathrm{d} \Omega ^2 \,,
\end{equation}
with $\mathrm{d} \Omega ^2= \mathrm{d} \theta ^2 + \sin^2 \theta \ \mathrm{d}  \phi ^2$ and
\begin{equation}
\label{Crbh_potential}
b^2(r) = 1-\frac{2m(r)}{r} =  \frac{1}{a^2(r)} \,.
\end{equation}
 Then we have to  solve the Einstein equations with a stress-energy tensor $T^{\mu}_{\nu}$ that can be specified from the Schwarzschild-like property $g_{tt}=-g_{rr}^{-1}\,$, as well as from the energy-momentum conservation $\nabla_{\mu} T^{\mu}_{\nu}=0\,$. With such conditions, the  tensor  $T^{\mu}_{\nu}$ has the following profile 
\begin{equation}
T^{\mu}_{\nu} = \mathrm{Diag} \big( - \rho(r)\,, p_r \,, p_{\perp} \,, p_{\perp} \big) \,,
\end{equation}
with
\begin{equation}
p_r = - \rho(r)  \qquad \mathrm{and} \qquad  p_{\perp} = -  \rho(r) - \frac{r}{2} \partial _r \rho(r) \,.
\end{equation}
Such a  tensor characterizes a self-gravitating droplet of anisotropic fluid with two pressure terms; one radial pressure $p_r$ and one tangential pressure $p_{\perp}\,$. On physical grounds, a non-vanishing radial pressure  balances the inward gravitational pull and prevents the source to collapse into a singular matter-point. It has also been shown that noncommutativity affects the matter sector \cite{MoN+10}. The geometry sector is modified accordingly, even if the Einstein tensor is formally kept unchanged.  By solving the 
Einstein equations one finds the following metric potential
\begin{equation} \label{Cncmetric}
b^2(r) = 1 - \frac{4M}{r \sqrt{\pi}} \gamma(3/2,r^2/4\theta) \,,
\end{equation}
where
\begin{equation}
\gamma(3/2,r^2/4\theta) = \int\limits_{0}^{r^2/4\theta} \mathrm{d}t \ t^{1/2} e^{-t}
\end{equation}
is the lower incomplete Gamma function. One can introduce a mass function for the above line element, namely  $m(r)= \frac{2M}{\sqrt{\pi}} \gamma(3/2,r^2/4\theta) \,$, that respects the Schwarzschild solution at large distances, since $\gamma(3/2,r^2/4\theta) |_{r^2 \gg 4\theta} \approx \sqrt{\pi}/2\,$.   
 Near the origin ($r \ll \sqrt{\theta}$) we can approximate $\gamma(3/2,r^2/4\theta)|_{r^2 \ll 4\theta} \approx \frac{r^3}{12 \theta^{3/2}}$ and  \eqref{Cncmetric} becomes
\begin{equation}
b^2(r) \approx 1 - \frac{\Lambda_{\mathrm{eff}}}{3}  r^2 \,,
\end{equation}
where $\Lambda_{\mathrm{eff}}= \frac{M}{\sqrt{\pi} \theta^{3/2}}\,$. Therefore, the singular origin of the Schwarzschild black hole has been replaced by a regular repulsive de Sitter-like core   with a constant and positive curvature, since the Ricci scalar is finite at the center and reads $R(0)=\frac{4M}{\sqrt{\pi} \theta^{3/2}}=4\Lambda_{\rm eff}\,$. Moreover, by solving the horizon equation  one finds two horizons (one Cauchy and one event horizon) for $M >M_0$ where $M_0 = 1.9 \sqrt{\theta}$. For $M=M_0$  one finds a degenerate horizon, while for $M<M_0$  no horizons exist. The energy scale $M_0$ can be seen as the transition point between black holes and elementary particle.

\begin{figure}[h!] 
\includegraphics[width=0.47 \textwidth]{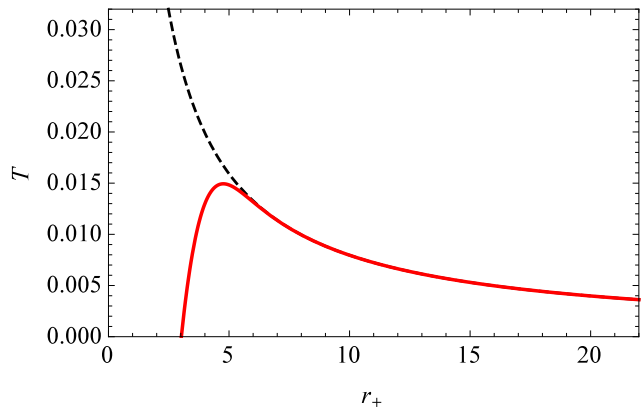}
\caption{The red solid curve represents the noncommutative black hole temperature $T$ \textit{vs} $r_+$  in $\sqrt{\theta}$ units, while the dashed black curve represents  the Hawking  temperature of a classical Schwarzschild black hole, appearing an ultraviolet divergence to the final stage of the evaporation.}
\label{fig:ncT}
\end{figure}

From a thermodynamic viewpoint, we have an improvement of the standard Hawking temperature. In this case, it is given by
\begin{equation} \label{Cnctemp}
T = \frac{1}{4\pi r_+} \left[ 1 - \frac{ r_+^3e^{-r_+^2/4\theta}}{4 \theta^{3/2} \gamma(3/2,r_+^2/4\theta)} \right] \,,
\end{equation}
with the event horizon $r_+$ satisfying the condition $b(r_+)=0\,$. In Fig.~\ref{fig:ncT} we plot the temperature \eqref{Cnctemp}.
In contrast to the classical case, the regular black hole temperature reaches a maximum before undergoing a positive heat capacity phase. Accordingly, the maximum of the temperature corresponds to a Hawking-Page-like transition between thermal radiation and a small stable black hole. The evaporation stops with  a zero temperature remnant configuration with radius $r_0=3\sqrt{\theta}\,$. According to this scenario, the quantum back reaction is strongly suppressed and it is possible to extend the validity of the semiclassical approximation to the final stage of the evaporation. The black hole actually cools down by emitting less and less energy. 
This concept is very similar to the known Rayleigh-Jeans catastrophe, where classical physics predicts  an ultraviolet divergence at short wavelengths of the black-body radiation. Quantum mechanics cures this pathology much in the same way noncommutativity cures the ultraviolet divergence of the  Hawking radiation.

\subsection{The black hole-cavity system}
\label{sub:bhc}

Following the seminal paper of Braden \textit{et al} \cite{BBW+90},  as well as similar works mentioned in the introduction, we have to take into account certain conditions that can be applied to any \textit{black hole-cavity} system: 
\begin{enumerate}
\item   We consider a spherically symmetric Euclidean line element of the form
\begin{equation} \label{Cbhcav_metric}
\mathrm{d} s^2 = b(r)^2 \mathrm{d} \tau^2 + a(r)^2 \mathrm{d} r^2 + r^2 \mathrm{d} \Omega ^2 \,,
\end{equation}
which is the same as in the aforementioned papers of the introduction but slightly simplified (this form has also been used in \cite{Lun08}). The Euclidean time $\tau$  comes after the analytical continuation $\tau=it\,$, providing us with a positive-definite Euclidean metric.
\item The event horizon $r_+$ is the inner boundary of the system with topology of a  two-sphere $S^2$ and  two constraints:
\begin{equation} \label{Cbound1}
b(r_+) = 0  \qquad \mathrm{and} \qquad \frac{b'(r)}{a(r)} \bigg |_{r+}=1 \,,
\end{equation}
where the prime symbol ($'$) denotes from now on differentiation with respect to $r\,$. The second constraint is a regularity condition, arising from the fact that we wish the geometry of the $\tau-r$ plane to look like a flat disk near $r_+\,$, in order to avoid conical singularities.
\item The cavity wall  is the outer boundary of the system with  topology  $S^1 \times S^2$. The wall is located at distance $r_\mathrm{c}$ and has an $S^2$ component with an area of $4\pi r^2_{\mathrm{c}}\,$. In addition,  heat can flow in either direction of the  wall  maintaining the temperature fixed. This way the black hole can be in thermal equilibrium with the cavity. The temperature of the boundary equals with the inverse of the period $\beta\,$, which is simply the proper length of the circle $S^1$ of the boundary:
\begin{equation} \label{Cbound2}
T^{-1} = \beta = \int\limits^{2\pi}_0 \mathrm{d}\tau \ b(r_{\mathrm{c}}) = 2\pi b(r_{\mathrm{c}}) \,.
\end{equation}
We have chosen the periodicity of $\tau$ to be $2\pi$ for convenience.
\item Applying the Gauss-Bonnet-Chern theorem, we find that the Euler number of the manifold should be $\chi =2\,$. That gives us one more constraint:
\begin{equation} \label{Cbound3}
\frac{1}{a^2(r)} \bigg|_{r_+} = 0 \,.
\end{equation}
\end{enumerate}
Now that we have given the above boundary conditions, we can proceed to the calculation of the reduced action for the desired system.

\section{The reduced action}
\label{sec:3}

The starting point is the Euclidean version of the known Einstein-Hilbert action with the  Gibbons-Hawking-York boundary term:
\begin{equation}
I=- \int\limits_{\mathcal{M}} \mathrm{d}^4 x \sqrt{g} \left( \frac{R}{16\pi} + \mathcal{L}_{\mathrm{m}} \right) - \frac{1}{8\pi} \int\limits_{\partial \mathcal{M}} \mathrm{d}^3 x \sqrt{h}  \big( K-K_0 \big)  \,,
\end{equation}
where $g$ and $h$ are the determinants of the Euclidean  metric of our manifold $\mathcal{M}$ and the induced metric on the boundary $\partial \mathcal{M}$ respectively, $R$ is the Ricci scalar, $\mathcal{L}_{\mathrm{m}}$ is the matter Lagrangian, $K$ is the trace of the extrinsic curvature on the boundary  and $K_0$ is a term   of our reference choice, analogous to $K\,,$ that is needed to regulate the boundary action. For notational convenience, we split the above action into 3 pieces (gravitational, matter and surface action)
\begin{eqnarray} \label{CI1}
I_{\mathrm{g}} &=& - \frac{1}{16\pi} \int\limits_{\mathcal{M}} \mathrm{d}^4 x \sqrt{g}  R \,, \\ \label{CI2}
I_{\mathrm{m}} &=& - \int\limits_{\mathcal{M}} \mathrm{d}^4 x \sqrt{g}  \mathcal{L}_{\mathrm{m}} \,, \\ \label{CI3}
I_{\mathrm{s}}  &=& -  \frac{1}{8\pi} \int\limits_{\partial \mathcal{M}} \mathrm{d}^3 x \sqrt{h} \big( K-K_0 \big) \,,
\end{eqnarray}
so that $I=I_{\mathrm{g}}+I_{\mathrm{m}}+I_{\mathrm{s}}\,$. The major difference from the classical system \textit{singular black hole-cavity}  is that the matter action \eqref{CI2}  is no longer zero for the regular case. We evaluate first the gravitational action $I_{\mathrm{g}}\,$. For a metric of the form \eqref{Cbhcav_metric}, the quantity $\sqrt{g} R$ is calculated to be
\begin{equation} \label{Cexpr1}
\sqrt{g} R = \left[ -\frac{2b}{a} + 2ba -\frac{4r b'}{a} - 4rb \left( \frac{1}{a} \right)'-2r^2 \left( \frac{b'}{a} \right)'   \right] \sin \theta \,,
\end{equation}
where $a=a(r)$ and $b=b(r)\,$. Plugging \eqref{Cexpr1} into the action \eqref{CI1} and integrating over the imaginary time and over the two angles $\theta$ and $\phi \,$, we get
\begin{equation}
I_{\mathrm{g}} =   \pi \int\limits_{r_+}^{r_{\mathrm{c}}} \mathrm{d}r \left(  \frac{1}{a} -a  - \frac{2r a'}{a^2} \right) b + \left(  \pi r^2 \frac{b'}{a}\right)  \bigg|_{r_{\mathrm{c}}} - \pi r_+^2 \,, 
\end{equation}
where we have performed integration by parts   and  used the boundary conditions \eqref{Cbound1}. Proceeding a bit further, we use the Hamiltonian constraints
\begin{equation}
G^{\tau}_{\tau} = G^{r}_{r}= -8 \pi \rho \,,
\end{equation}
where $\rho=\rho(r)$ is the matter density profile of the regular black hole. The components of the Einstein tensor read
\begin{equation}
G^{\tau}_{\tau} = \frac{ \left( a - a^3 - 2r a' \right)}{r^2 a^3}  \qquad \mathrm{and} \qquad  G^{r}_{r} = \frac{b-a^2b +2r b'}{r^2 a^2 b} \,.
\end{equation}
From the relation $G^{\tau}_{\tau} / G^{r}_{r} =1\,$, we conclude to $b=\mathcal{C}/a\,$, where the integration constant $\mathcal{C}$ can be specified, if the profile $\rho(r)$ of the regular black hole is specified. For now we can use the condition \eqref{Cbound2}  to write $\mathcal{C}=\frac{\beta a(r_\mathrm{c})}{2\pi}\,$. Under these circumstances, the gravitational action can be written as
\begin{equation}
I_{\mathrm{g}} =  \frac{\beta a(r_\mathrm{c})}{2} \left[ r  \left( \frac{1}{a^2}-1 \right) \right] \bigg|_{r_+}^{r_{\mathrm{c}}} - \frac{\beta }{2} \left(   \frac{r^2 a'}{a^2}\right)  \bigg|_{r_{\mathrm{c}}} - \pi r_+^2 \,.
\end{equation}
 Next we shift our interest to the matter action \eqref{CI2}. Following the reasoning of  \cite{MaN11}, our effective fluid-like approach takes leading order quantum geometry effects, while letting us formally work in a classical framework. This allows us to connect our matter Lagrangian with the "on shell" action of an anisotropic fluid that reads
\begin{equation}
\mathcal{S}(\mathrm{on \ shell}) = \int \mathrm{d}^4x \sqrt{-g} \big( 2p_{\mathrm{r}} - p_{\perp}  \big) \,,
\end{equation}
identifying our matter Lagrangian with
\begin{equation} \label{Cmatter_Lagr}
\mathcal{L}_{\mathrm{m}} =  2p_{\mathrm{r}} - p_{\perp} = -\rho + \frac{r}{2} \rho'  \,.
\end{equation}
Plugging \eqref{Cmatter_Lagr} into the action \eqref{CI2} and performing the integration, we get
\begin{equation} \label{Im_middle}
I_{\mathrm{m}} = 8\pi^2 \int\limits_{r_+}^{r_{\mathrm{c}}} \mathrm{d}r \ abr^2 (\rho-\frac{r}{2} \rho') \,.
\end{equation}
From the Hamiltonian constraint $G^{\tau}_{\tau} =\frac{1}{r^2} \left[ r \left( \frac{1}{a^2}-1 \right)  \right] '= -8 \pi \rho\,$, we can extract an expression for the integral $\int\limits_{r_+}^{r_{\mathrm{c}}} \mathrm{d}r \ r^2 \rho \,$:
\begin{equation} \label{C_expr1}
\int\limits_{r_+}^{r_{\mathrm{c}}} \mathrm{d}r \ r^2 \rho = -\frac{1}{8\pi} \left[ r \left( \frac{1}{a^2}-1 \right)  \right] \bigg|_{r_+}^{r_{\mathrm{c}}} \,.
\end{equation}
Thus using \eqref{C_expr1}, the relation $ab=\mathcal{C}=\frac{\beta a(r_\mathrm{c})}{2\pi}\,$, and integrating by parts the $\rho'$-term in \eqref{Im_middle}, we conclude to 
\begin{equation} \label{Im_final}
I_{\mathrm{m}} = 4 \pi \beta a(r_{\mathrm{c}}) \left[ \frac{5}{16\pi}  \left( r-\frac{r}{a^2} \right) \bigg|_{r_+}^{r_{\mathrm{c}}} - \frac{1}{2} r^3 \rho \Big|_{r_+}^{r_{\mathrm{c}}} \right] \,.
\end{equation}
We continue with the derivation of the surface action \eqref{CI3}. Our boundary is a timelike surface at $r_{\mathrm{c}}=\mathrm{const.}\,$, described by a radial spacelike normal unit vector $\eta^{\mu}=\frac{\delta^{\mu}_{r}}{\sqrt{g_{rr}}}\,$ ($\eta^{\mu}\eta_{\mu}=1$) perpendicular to the boundary. Thus the induced metric on the boundary can be written as $h_{\mu\nu}=g_{\mu\nu}-\eta_{\mu}\eta_{\nu}$ and is described by a line element of the form
\begin{equation}
\mathrm{d} s ^2 = b(r_{\mathrm{c}})^2 \mathrm{d} \tau ^2 + r_{\mathrm{c}}^2 \mathrm{d} \Omega ^2 \,.
\end{equation}
This allows us to define the trace of the extrinsic curvature $K$ as the divergence of the normal unit vector $\eta^{\mu}\,$, giving
\begin{equation} \label{CK}
K=\nabla_{\mu} \eta ^{\mu} = \frac{\left( br^2 \right)' }{abr^2} \bigg|_{r_{\mathrm{c}}} \,.
\end{equation}
As for $K_0\,$, we can choose any space we wish as the reference point from which the energy is measured. In our case, we choose $K_0$ such that $I=0$ when $M=0\,,$ corresponding to an asymptotically flat spacetime that gives
\begin{equation} \label{CK0}
K_0 = \frac{2}{r_{\mathrm{c}}} \,.
\end{equation}
Using  the above expressions for $K$ and $K_0\,$, 
the surface action  becomes
\begin{equation}
I_{\mathrm{s}} = \beta \left( \frac{r^2 a'}{2a^2}  - \frac{r}{a} +r \right) \bigg|_{r_{\mathrm{c}}} \,.
\end{equation}
Finally, from the sum $I_{\mathrm{g}}+I_{\mathrm{m}}+I_{\mathrm{s}}$ we obtain the reduced action $I$ that reads
\begin{equation} \label{CIrbh_cav}
I= \beta \left\lbrace   -  \frac{a(r_{\mathrm{c}})}{2}    \left(r - \frac{r}{a^2} \right) \bigg|_{r_+}^{r_{\mathrm{c}}} + 4\pi a(r_{\mathrm{c}}) \left[ \frac{5}{16\pi}  \left( r-\frac{r}{a^2} \right) \bigg|_{r_+}^{r_{\mathrm{c}}} - \frac{1}{2} r^3 \rho \Big|_{r_+}^{r_{\mathrm{c}}} \right] - \frac{r_{\mathrm{c}}}{a(r_{\mathrm{c}})}  + r_{\mathrm{c}}        \right\rbrace - \pi r_+^2 \,.
\end{equation}
For large distances relative to the minimal cut-off length, the mass density profile of the black hole becomes proportional to the Dirac delta function, i.e., $\rho(r) \sim M \delta(r)\,$, meaning that $\rho(r_+) = \rho(r_{\mathrm{c}} ) = 0$ since $r_+,r_{\mathrm{c}} \neq 0\,$. Then the Hamiltonian constraints give   $\frac{1}{a(r)^2} \approx 1- \frac{r_+}{r}\,$, implying the relation $\left( \frac{r}{a^2}-r \right) \Big|_{r_+}^{r_{\mathrm{c}}} \approx 0$. Therefore, in this limit  the matter action \eqref{Im_final} vanishes and the reduced action \eqref{CIrbh_cav}  reduces to the action of the classical uncharged system \textit{singular black hole-cavity} \cite{Lun08}:
\begin{equation}
I \approx \beta r_{\mathrm{c}}  \left( 1 - \sqrt{1-\frac{r_+}{ r_{\mathrm{c}}}} \right) - \pi r_+^2 \,.
\end{equation}

\section{The phase structure}
\label{sec:4}
\begin{figure}[h!] 
\includegraphics[width=0.47 \textwidth]{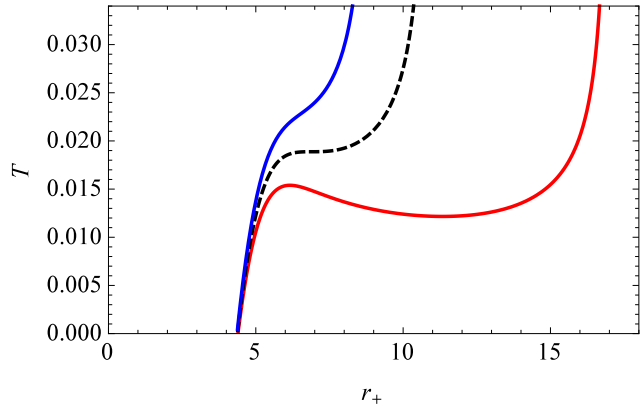}
\caption{The noncommutative black hole temperature $T$ inside the cavity \textit{vs} $r_+$ in $\sqrt{\theta}$ units, for $r_{\mathrm{c}}=17 \sqrt{\theta}$ (red solid curve), for $r_{\mathrm{c}} = r_{\mathrm{i}} \simeq 10.913 \sqrt{\theta}$ (black dashed curve) and for $r_{\mathrm{c}}=9 \sqrt{\theta}$ (blue solid curve).}
\label{fig:CTcav}
\end{figure}

\begin{figure}[h!] 
\includegraphics[width=0.47 \textwidth]{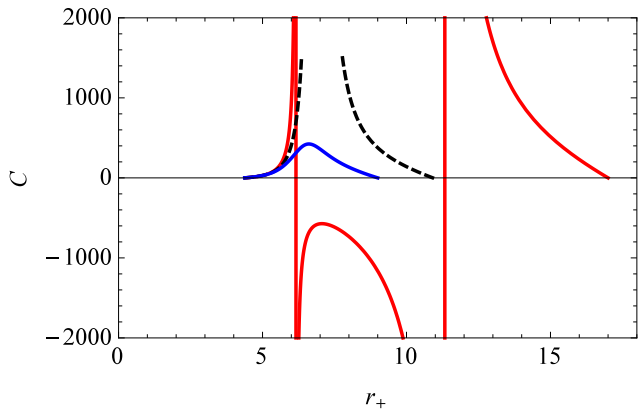}
\caption{The heat capacity $C$ of the noncommutative black hole inside the cavity \textit{vs} $r_+$ in $\sqrt{\theta}$ units, for $r_{\mathrm{c}}=17 \sqrt{\theta}$ (red solid curve), for $r_{\mathrm{c}} = r_{\mathrm{i}} \simeq 10.913 \sqrt{\theta}$ (black dashed curve) and for $r_{\mathrm{c}}=9 \sqrt{\theta}$ (blue solid curve).}
\label{fig:CCcav}
\end{figure}

In this last section we apply the reduced action formalism to the case of a noncommutative Schwarzschild black hole as an example. 
Then we derive the desired thermodynamic quantities  and we study possible phase transitions that take place inside the cavity.		
For the noncommutative case, the profile of the density is given by  \eqref{Cncdensity}. Solving the Hamiltonian constraints for this profile,  the metric functions $a$ and $b$ of the line element \eqref{Cbhcav_metric} are evaluated to be
\begin{equation}
a(r)= \frac{1}{\sqrt{1 - \frac{4M}{r \sqrt{\pi}} \gamma(r)}} \qquad \mathrm{and} \qquad b(r)= \mathcal{C} / a(r) = \frac{\beta}{2\pi \sqrt{1-\frac{4M}{r_{\mathrm{c}} \sqrt{\pi}} \gamma(r_{\mathrm{c}})}}   \sqrt{1-\frac{4M}{r \sqrt{\pi}} \gamma(r)} \,,
\end{equation}
where $\gamma(r)=\gamma(3/2,r^2/4\theta) \,$.
 Moreover, the boundary condition \eqref{Cbound3} implies the relation $M=\frac{r_+ \sqrt{\pi}}{4 \gamma(r_+)}\,$. Plugging these expressions into the general form of the reduced action \eqref{CIrbh_cav}, we get the form of the noncommutative reduced action $I_{\ast}$:
\begin{eqnarray} \nonumber
I_{\ast} &=& \beta  r_{\mathrm{c}} \left(  1 - \sqrt{1-\frac{r_+ \gamma (r_{\mathrm{c}})}{r_{\mathrm{c}} \gamma (r_+)}} \right) + \frac{\beta r_+}{2} \left(  \frac{1-\gamma(r_{\mathrm{c}})/\gamma (r_+)}{\sqrt{1-\frac{r_+ \gamma(r_{\mathrm{c}})}{r_{\mathrm{c}} \gamma(r_+)}}} \right) \\
&& + \frac{4\pi \beta}{\sqrt{1-\frac{r_+ \gamma(r_{\mathrm{c}})}{r_{\mathrm{c}} \gamma(r_+)}}} \left[  \frac{5 r_+}{16\pi} \left( \frac{ \gamma(r_{\mathrm{c}})}{\gamma(r_+)} -1 \right) - \frac{1}{2} r_{\mathrm{c}}^3 \rho(r_\mathrm{c})  + \frac{1}{2} r_+^3 \rho(r_+)  \right] - \pi r_+^2 \,.
\end{eqnarray}
As discussed in \cite{BrY94}, thermodynamic quantities can be derived from the reduced action. The canonical partition function $Z\,$, evaluated in the zero-loop approximation, is connected to the Euclidean action  through the relation $Z = e^{-\beta F} \approx e^{-I_{\ast}}\,$, where $F$ is the Helmholtz free energy. 
The inverse temperature  can be derived by extremizing the action with respect to $r_+$ and solving for $\beta\,$:
\begin{equation} \label{Ctemp_beta}
\frac{\partial I_{\ast}(\beta,\theta,r_+,r_{\mathrm{c}})}{\partial r_+}=0 \qquad \longrightarrow \qquad \beta=\beta(\theta,r_+,r_{\mathrm{c}})=T^{-1} \,.
\end{equation}	
Then the entropy of the system can be found from the relation
\begin{equation}
S = \beta \left( \frac{\partial I_{\ast}}{\partial \beta} \right)_{\theta,r_{\rm c}} - I_{\ast} = \beta \left( \frac{\partial I_{\ast}}{\partial r_+} \right)_{\theta,r_{\rm c}} \left( \frac{\partial \beta}{\partial r_+} \right)^{-1}_{\theta,r_{\rm c}} - I_{\ast} = \pi r_+^2 \,,
\end{equation}
giving the known area law. The  expression of the temperature is omitted here since it is rather lengthy but its plot is illustrated in Fig.~\ref{fig:CTcav} for different horizons inside the cavity. 
Depending on the fixed distance $r_{\mathrm{c}}$ of the cavity wall relative to the minimal length $\sqrt{\theta}\,$, the temperature may appear two  (red curve in Fig.~\ref{fig:CTcav}), one  (black dashed curve in Fig.~\ref{fig:CTcav}) or no extrema (blue curve in Fig.~\ref{fig:CTcav}).  For $r_{\mathrm{c}} = r_{\mathrm{i}}$ the two extrema of the temperature merge at one inflexion point, while for $r_{\mathrm{c}} > r_{\mathrm{i}}$ the system  appears always a local maximum and a local minimum temperature. The cavity has already changed the picture relative to the pure asymptotically flat regular case \eqref{Cnctemp}, where there is only one extremum for the temperature, as can be seen in Fig~\ref{fig:ncT}. Inside the cavity there is a strong indication of a different phase transition when  $r_{\mathrm{c}} \geq r_{\mathrm{i}}\,$. This can be checked by calculating the heat capacity 
\begin{equation}
C=T \left(  \frac{\partial S}{\partial T} \right)  = - \beta^2 \left(  \frac{\partial^2 I}{\partial \beta ^2} \right)_{\theta,r_{\rm c}}  \,,
\end{equation}
 whose form is also omitted here  but its shape is plotted in Fig.~\ref{fig:CCcav}.
For $r_{\mathrm{c}} > r_{\mathrm{i}} $ (red curve  in Fig.~\ref{fig:CCcav}) the heat capacity appears two divergent regions where $\mathrm{d} T=0\,$. In this case, there exists a small locally stable black hole with $C>0$ when $r_+$ is smaller than the first divergent region. Then it turns to a locally unstable intermediate black hole with $C<0$ when $r_+$ lies between the two divergences and finally it turns into a locally  stable large black hole with $C>0$ when $r_+$ is bigger than the second divergent region. Hence, a phase transition seems to take place between a small and a large stable black hole inside the cavity, in contrast to the asymptotically flat case of Sec.~\ref{sec:2} without a cavity, where the transition occurs between background radiation and a small stable black hole. This small/large stable black hole transition seems to be similar with  the  liquid/gas transition of a Van der Waals gas \cite{KuM12}, in contrast to the Hawking-Page-like transition of the \textit{singular black hole-cavity} system \cite{Lun08}. For $r_{\mathrm{c}} = r_{\mathrm{i}} $ (black dashed curve in Fig.~\ref{fig:CCcav}) the two stable black holes coexist at one inflexion point, while for $r_{\mathrm{c}} < r_{\mathrm{i}} $ (blue curve in Fig.~\ref{fig:CCcav}) there exists a single thermally stable black hole with no phase transitions and with a temperature which is monotonically increasing with  the horizon.

Similar phase structure appear also   regular  black holes \cite{NiT11,Tzi19} 
 inside an AdS background  without a cavity. In other words, either confining the regular black hole in a cavity, or placing it inside an AdS space, the black hole appears a small/large black hole phase transition, mimicking the liquid/gas transition of a Van der Waals gas. This indicates that AdS holography might not depend on some properties of AdS spacetime, but rather on the confinement characteristic, as has also been noticed in \cite{Lun08}.

\begin{figure}[h!] 
\includegraphics[width=0.47 \textwidth]{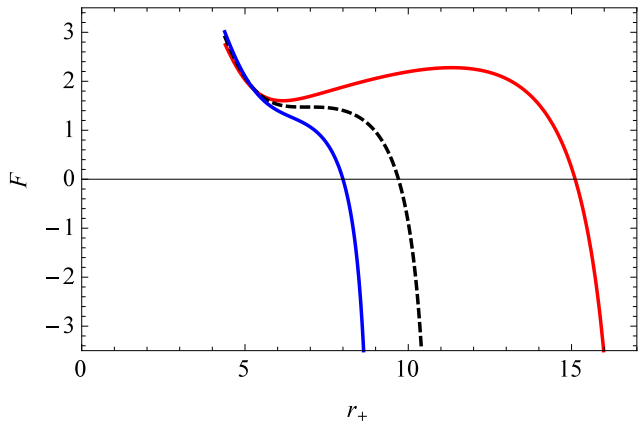}
\caption{The free energy $F$ of the noncommutative black hole inside the cavity \textit{vs} $r_+$ in $\sqrt{\theta}$ units, for $r_{\mathrm{c}}=17 \sqrt{\theta}$ (red solid curve), for $r_{\mathrm{c}} = r_{\mathrm{i}} \simeq 10.913 \sqrt{\theta}$ (black dashed curve) and for $r_{\mathrm{c}}=9 \sqrt{\theta}$ (blue solid curve).}
\label{fig:CFcav}
\end{figure}

Last but not least, the Helmholtz free energy $F$ is the thermodynamic potential governing our system and is identified to be $ F= I_{\ast} / \beta \,$.
Its form is illustrated in Fig.~\ref{fig:CFcav}. For $r_{\mathrm{c}} > r_{\mathrm{i}}$ the free energy appears two extrema (red curve in Fig.~\ref{fig:CFcav}) with an unstable middle branch of an increasing $F$ due to the phase transition. These extrema correspond to the two divergences of the heat capacity, as can be seen from the above plots. For $r_{\mathrm{c}} < r_{\mathrm{i}}$ the free energy is a monotonically decreasing function of $r_+$ (blue curve in Fig.~\ref{fig:CFcav}) with no indication of a phase transition, signalling this way the local and global stability of the system. For $r_{\mathrm{c}} = r_{\mathrm{i}}$ the two extrema of $F$ merge at one inflexion point (black dashed curve in Fig.~\ref{fig:CFcav}).

\section{Conclusions}
\label{sec:concl}
We have presented a geometric and a thermodynamic analysis of a regular black hole enclosed by a cavity. The temperature at the walls and the radius of the cavity are held fixed, allowing us to work in a canonical ensemble. Based on the assumption that any  regular black hole solution is described by a source of anisotropic fluid with two pressure terms, we performed the derivation of the reduced action. The new action is a generalization of the one describing a singular black hole, since it contains an extra matter action-piece that vanishes in the classical limit. Applying this formalism to the profile of the noncommutative case, we derived through the reduced  action all the necessary thermodynamic quantities, in order to study the phase structure of the system. The conclusion is that, above a certain value of the position of the cavity wall, i.e., $r_{\mathrm{c}} \geq r_{\rm i}\,$, the  black hole can undergo  a phase transition between a small/large stable black hole, similar to the liquid/gas transition of a Van der Waals gas,  while for  $r_{\mathrm{c}} < r_{\rm i}$ there exists always a thermally stable remnant. The enclosing cavity provides similar phase structure  with the system of a  regular  black hole placed in AdS space without a cavity.    

\acknowledgements
The present work has been supported by the GRADE Completion Scholarships, which are funded by the STIBET program of the German Academic Exchange Service (DAAD) and the Stiftung zur F\"orderung der internationalen wissenschaftlichen Beziehungen der Johann Wolfgang Goethe-Universit\"at. The author would also like to thank Piero Nicolini  for reading carefully the manuscript and for providing useful comments.

\end{document}